# Field-free spin-orbit torque switching through domain wall motion

Neil Murray[1], Wei-Bang Liao[1], Ting-Chien Wang[1], Liang-Juan Chang[2], Li-Zai Tsai[2], Tsung-Yu Tsai[1], Shang-Fan Lee[2], and Chi-Feng Pai[1,3*]

[1]*Department of Materials Science and Engineering, National Taiwan University, Taipei 10617, Taiwan*
[2]*Institute of Physics, Academia Sinica, Taipei 10617, Taiwan*
[3]*Center of Atomic Initiative for New Materials, National Taiwan University, Taipei 10617, Taiwan*

Deterministic current-induced spin-orbit torque (SOT) switching of magnetization in a heavy transition metal/ferromagnetic metal/oxide magnetic heterostructure with the ferromagnetic layer being perpendicularly-magnetized typically requires an externally-applied in-plane field to break the switching symmetry. We show that by inserting an in-plane magnetized ferromagnetic layer CoFeB underneath the conventional W/CoFeB/MgO SOT heterostructure, deterministic SOT switching of the perpendicularly-magnetized top CoFeB layer can be realized without the need of in-plane bias field. Kerr imaging study further unveils that the observed switching is mainly dominated by domain nucleation and domain wall motion, which might limit the potentiality of using this type of multilayer stack design for nanoscale SOT-MRAM application. Comparison of the experimental switching behavior with micromagnetic simulations reveals that the deterministic switching in our devices cannot be explained by the stray field contribution of the in-plane magnetized layer, and the roughness-caused Néel coupling effect might play a more important role in achieving the observed field-free deterministic switching.

---

[*] Email: cfpai@ntu.edu.tw

## I. INTRODUCTION

Current-induced spin-orbital effects in heavy transition metals (HMs) can be utilized to generate spin-orbit torques (SOTs) acting on adjacent ferromagnetic (FM) layers [1,2]. However, if the FM layer has perpendicular magnetic anisotropy (PMA), an externally applied in-plane field is generally necessary in order to achieve deterministic current-induced SOT switching. This in-plane bias field serves as either a symmetry-breaking role for macrospin scenario [3] or the external force to re-align chiral domain wall moments in the magnetic layer for (multi-) domain nucleation-propagation scenario [4]. Since state-of-the-art magnetic random access memory (MRAM) designs mostly consist of FM layers with PMA [5], it is therefore crucial to eliminate the need of external field for constructing a truly useful SOT-MRAM device. Several pioneering works have shown that deterministic and/or tentative field-free SOT switching can be realized by introducing exchange bias field that originates from an antiferromagnetic (AFM) layer such as IrMn [6-9], which are typically grown in adjacent to the FM layers to be switched. The deterministic switching can also be achieved by inserting an extra FM layer into the conventional HM/FM/oxide heterostructure, becoming a FM(in-plane)/HM/FM(PMA)/oxide structure, which is more compatible with contemporary perpendicular MRAM architecture while compared to the AFM approach [10,11]. Other recent approaches include the employment of lateral wedge structure [12,13], geometrical engineering [14-16], ferroelectric control [17], and countering spin current buffer layer [18].

In this work, we show that insertion of an extra in-plane magnetized FM layer allows for deterministic current-induced SOT switching of the PMA FM layer without the need of an in-plane biasing field and investigate the possible influence from the in-plane FM layer effective field. Utilizing MOKE imaging we further show that switching is mainly driven by domain wall nucleation and domain wall propagation, with the polarity of switching loops being determined by the in-plane layer magnetization direction. Even by placing an insulating insertion layer in between the in-plane FM layer and the HM layer to block potential spin current from the in-plane FM layer, field-free switching is still achievable, which suggests that the observed phenomenon is governed by a field effect from the in-plane FM layer. However, micromagnetic simulations of the experimental structure show that in an ideal case, the polarity of the switching loop should be determined by the direction of the stray field, which is opposite to the direction of the in-plane FM layer magnetization. Based on these observations, we conclude that the deterministic SOT switching polarity in our samples is primarily determined by the effective field from the Néel orange-peel effect.

## II. SAMPLE PREPARATION

All samples were deposited in multilayer stacks on thermally oxidized Si substrates via DC/RF magnetron sputtering at room temperature. The magnetron sputtering system was kept at a base pressure of $3 \times 10^{-8}$ Torr and an Ar working pressure of 3 mTorr and 10 mTorr for DC and RF sputtering, respectively. For metallic materials a power of 30 W was used for deposition, while for the insulating layer a power of 50 W was used. Two groups of multilayers were prepared as shown in Fig. 1: (a) A control series of W($t_W$)/CoFeB(1.4)/MgO(1.6)/Ta(2) (numbers in parenthesis are layer thickness in nanometers), and (b) the experimental series CoFeB(3)/W($t_W$)/CoFeB(1.4)/MgO(1.6)/Ta(2). W thickness $t_W$ were set to be 1, 2, and 3 nm and the CoFeB target has a composition of $Co_{40}Fe_{40}B_{20}$. After deposition, the samples were annealed at 300 °C for one hour without applying magnetic field. For samples used in electrical measurements, Hall-bar devices with channel widths of 5 μm were fabricated using standard photolithography and subsequent lift-off processes. Following annealing, we confirmed our samples having PMA either by magneto-optical Kerr imaging (MOKE) or anomalous Hall effect (AHE) measurements under varying applied out-of-plane field strengths. The resulting square hysteresis loops (as seen in Fig.

1(c)) indicate that the top CoFeB(1.4) layer indeed has good PMA.

## III. EXPERIMENTAL RESULTS

### A. Current-induced SOT switching

As schematically shown in Fig. 2(a), we first performed conventional current-induced SOT switching tests on the experimental samples by passing 0.1-second current pulses of varying strength through the current channel of the Hall-bar device under different applied in-plane field strengths ($H_x$). The magnetization switching behavior was then detected by the anomalous Hall resistance $R_H$ variation of the top CoFeB(1.4) layer. To verify the switching is full or partial, we checked the field-swept $\Delta R_H \approx 6.5\ \Omega$, as shown in Fig. 2(b). We found that for all our Hall-bar samples, current-induced switching was possible even at low in-plane field strengths ($H_x \sim 50$ Oe). Furthermore, the SOT switching polarity was able to be controlled by the direction of applied in-plane field, as shown in Fig. 2(c) and (d) for the experimental series, which we attribute to the direction change of chiral domain wall (DW) moment in the top CoFeB layer [19,20]. It is noted that $\Delta R_H \approx 5.5\ \Omega$ was observed for the current-induced SOT switching measurements since only the region with current flowing through would be switched by SOT [13,21], as indicated in Fig. 2(a).

For field-free current-induced switching measurements, we applied an in-plane field for a short duration (3 seconds) before moving the device under test (experimental samples) to a location with no remnant fields from the electromagnet. Again by applying 0.1-second current pulses we were able to see nearly full deterministic switching results ($\Delta R_H \approx 5.5\ \Omega$), as seen in Fig. 3. It is also noted that even after being removed from the applied field, the switching polarity of this experimental sample was still determined by the direction of the pre-applied field. Following successful field-free switching, we preformed switching measurements at various consecutive low in-plane fields ($|H_x| \leq 20\,\text{Oe}$) after applying a large in-plane field. For a pre-magnetized field of $H_x = -500\,\text{Oe}$, as shown in Fig. 4(a), full switching can be achieved until a positive field of $H_x = 3\,\text{Oe}$ was applied. Similarly, when the sample was pre-magnetized with a positive in-plane field $H_x = 500\,\text{Oe}$, we did not see a full collapse of switching loop until $H_x$ reached -3 Oe (Fig. 4(b)). For both cases, application of a field greater than 3 Oe in the opposite direction of the pre-applied field caused the polarity of the switching loop to change. These results suggest the existence of a built-in symmetry breaking field of ~ 3 Oe, whose direction is parallel to the bottom in-plane CoFeB(3) magnetization direction.

In contrast, we were unable to observe deterministic switching from our control samples in the field-free testing (representative data in supplementary materials). In fact, even for the experimental series, namely CoFeB(3)/W($t_W$)/CoFeB(1.4)/MgO(1.6)/Ta(2), only devices made from the $t_W = 1$ nm film show signs of field-free switching. No conclusive deterministic switching can be found in samples with $t_W \geq 2\,\text{nm}$. This observation is quite different from a recent report by Chen *et. al.* [11], in which the W layer of their devices can be as thick as 7 nm and still show deterministic current-induced switching.

### B. MOKE imaging of SOT-induced switching

To further clarify the driving mechanism behind our observed field-free current-induced switching, we employed wide-field MOKE imaging approach to see the SOT-driven switching process within current channel. MOKE imaging revealed that when the pulsing current reaches a critical value, domain nucleation will occur at the channel edges followed by current-induced DW propagation (presumably driven by SOT). As shown in Fig. 3 alongside the switching $R_H$ data, in general, the DW will propagate from one end of the channel to the other and stop at the junction of current channel and electrode. Reversible switching can also be detected as the reversed DW motion while applying opposite currents (videos of these processes can be found in the online supplementary materials). Also note that magnetization switching, or to be more accurate, current-induced SOT-driven DW motion, occurs only in the current channel since the current density is higher in that region [21]. We also did not observe a systematic influence from the stripe domains outside the Hall-bar device affecting the domain wall propagation across the channel. More importantly, reversing the pre-applied magnetic field direction will also reverse the current-induced DW motion direction [19,20]. Our observation therefore suggests that the bottom CoFeB(3) layer can assist the formation of stable chiral domain wall with preferred orientation in the top CoFeB(1.4) layer, presumably through the Néel orange-peel effect [11]. However, note that our experimentally determined magnitude of the Néel effective field from the switching measurements (~ 3 Oe) is much smaller than that estimated by Chen *et. al*. in Ref. [11] (can reach as high as 15 Oe with 3.6 nm of W), which may be due to a smaller film roughness in our samples.

### C. Effective field or torque?

It is also important to address that other mechanisms such as exchange coupling of the FM layers with a W spacer layer or the recently proposed non-conventional spin currents generated by FM layers that are close to each other [10,22] can lead to a similar field-free switching result through an additional spin torque effect. Note that the thin W layer in our samples, which is much thinner than the reported 3.5 nm spin diffusion length of β-W [23], may well allow for a spin-filtering torque from the interface of the in-plane CoFeB layer and the W spacer layer [10] causing an additional torque on the PMA CoFeB layer through spin diffusion. In order to rule out this possibility, we performed the same current-induced switching measurements on samples with an extra MgO(1) insulating layer placed in between the CoFeB(3) layer and the W(1) layer, to block the transmission of spin current from the bottom in-plane CoFeB layer and/or the CoFeB(in-

plane)/W interface. Representative results from a CoFeB(3)/MgO(1)/W(1)/CoFeB(1.4)/MgO(2) device is shown in Fig. 4(c). It can be seen that even with the existence of an effective spin current transmission barrier MgO [24], current-induced field-free switching is still achievable. Therefore, we tentatively rule out the possible contribution of non-conventional spin current and spin torque effects from the in-plane CoFeB layer in our observed field-free switching phenomenon.

## IV. MICROMAGNETIC SIMULATIONS

Next, to show the bottom in-plane magnetized CoFeB layer's effect on the switching dynamics of the top CoFeB layer with PMA, we performed micromagnetic simulations using mumax$^3$ [25]. We modeled both the control group (W/CoFeB/MgO) and the experimental group (CoFeB/W/CoFeB/MgO) of devices. The simulation parameters were modeled after ideal parameters for CoFeB thin films (exact parameters in supplementary materials). Without the application of an external field, no current-induced switching was observed in the simulation of the control group. However, as shown in Fig. 5(a), with the application of an in-plane field along $x$-direction greater than 100 Oe (0.01 T) deterministic switching was observed, which agrees well with similar micromagnetic studies [26] and our own measured results. The switching polarity of current-induced switching loop in the control device was determined by the applied external field, which is also consistent with the measured results from W/CoFeB/MgO devices.

Simulations of the experimental device structure with an extra in-plane magnetized CoFeB layer showed full switching was possible with no applied field (Fig. 5(a)). The strength of the stray field in our simulated samples was also found to reach ~ 600 Oe (0.06 T) in the direction opposite to the in-plane CoFeB layer magnetization, as seen in Fig. 5(b). These results suggest that the stray field from the in-plane CoFeB layer can possibly provide a strong enough biasing field to enable field-free switching [27]. However, contrary to our experimental results, the polarity of the simulated current-induced SOT switching loops were opposite to what we measured (Fig. 3).

Moreover, the device structure modeled here does not account for magnetic films extending under the electrodes as in the experimental devices. In order for the stray field to provide a strong enough in-plane field to allow for domain nucleation and expansion, the magnetic film must have an edge at the hall bar. Therefore, the stray field contribution cannot successfully explain the field-free switching that we experimentally observed. Again, we tentatively conclude that the roughness-caused Néel orange-peel effect is more likely the dominating mechanism behind the observed field-free switching, which cannot be captured by the simulation due to the ideally flat simulated layer structures.

Nevertheless, to study the feasibility of employing FM(in-plane)/HM/FM(PMA) layer structure with a built-in in-plane field (either from the stray field or from the Néel field) in potential nano-sized SOT-MRAM devices, we further performed simulations on devices with elliptical-shaped geometries. In Fig. 6(a) we show a theoretical nano-pillar device structure where the pillar is on an extended layer of CoFeB. The in-plane magnetized CoFeB can be tentatively formed by over-etching the extended CoFeB layer to gain shape anisotropy along *x*-direction. Micromagnetic simulations of nano-pillar ellipses down to 10 nm in width (Fig 6(b)) show that deterministic switching is possible from the stray field contribution alone, as shown in Fig. 6(c). We also find that the critical switching current density increases with decreasing the device size (see Table 1). This result is in line with other previous works [28,29] which show that nanoscale devices tend towards coherent switching, with the critical current density increasing by an order of magnitude as domain nucleation becomes more difficult [30]. Also note that devices smaller than 20 nm by 10 nm no longer show deterministic switching, which might be a major obstacle of employing this stray field approach for SOT-MRAM devices at nanoscale. A representative current-induce SOT switching profile from a 10 nm by 5 nm nano-pillar is shown in Fig. 6(d), which indicates that the switching mode is stochastic rather than deterministic.

## V. CONCLUSION

To summarize, we experimentally demonstrated deterministic current-induced SOT switching in a CoFeB(in-plane magnetized)/W/CoFeB(PMA)/MgO magnetic heterostructure. Through MOKE imaging, we found the switching is mainly driven by domain nucleation and expansion across the current channel. The in-plane magnetized CoFeB layer provides a symmetry breaking in-plane field of ~ 3 Oe to facilitate field-free switching, whose origin is tentatively attributed to the Néel orange peel effect. Micromagnetic simulation confirms that the switching is mainly governed by domain nucleation and domain wall motion, with the polarity of such deterministic field-free SOT switching determined by the direction of stray field from the in-plane CoFeB layer. However, this polarity is opposite to our experimental observation. Therefore, we again conclude that the field-free switching observed in our devices should be explained by a Néel effective field scenario. We further shows via simulations that the switching mode of devices with built-in in-plane fields will change from deterministic to stochastic if the lateral size of the simulated device becomes less than ~ 10 nm, which might limit the feasibility of employing such FM(in-plane)/HM/FM(PMA) layer design for realistic SOT-MRAM applications.

## ACKNOWLEDGEMENTS

This work is supported by the Ministry of Science and Technology of Taiwan (MOST) under Grant No. 105-2112-M-002-007-MY3 and 108-2636-M-002-010 and by the Center of Atomic Initiative for New Materials (AI-Mat), National Taiwan University, Taipei, Taiwan from the Featured Areas Research Center Program within the framework of the Higher Education Sprout Project by the Ministry of Education (MOE) in Taiwan under grant No. NTU-107L9008. This work is also partly supported by Academia Sinica Thematic Project under Grant No. AS-107-TP-A04.

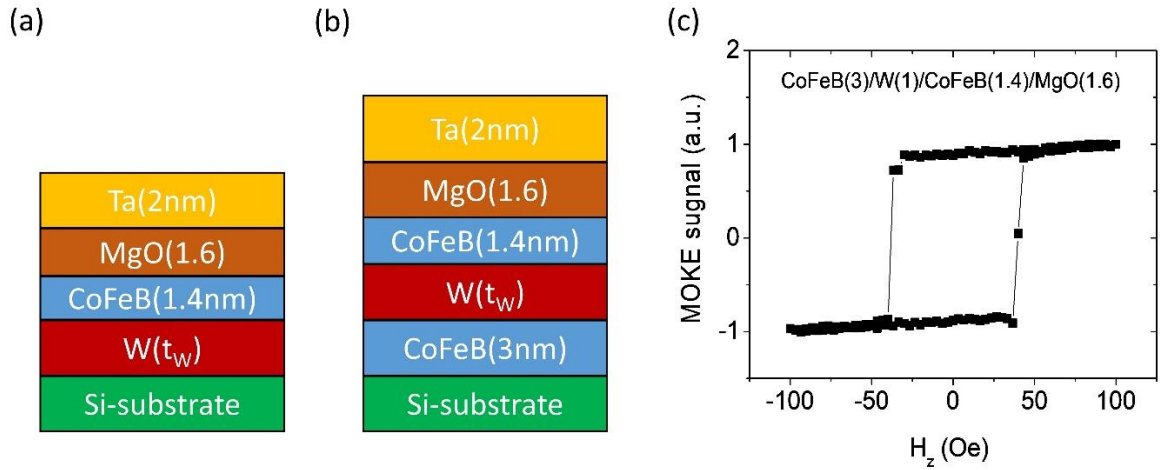

Figure 1. Multilayer stacks prepared for the present study: (a) W-based control sample, (b) experimental samples with the additional in-plane magnetized bottom CoFeB(3). (c) Representative out-of-plane hysteresis loop of a CoFeB(3)/W(1)/CoFeB(1.4)/MgO(1.6) thin film obtained by MOKE.

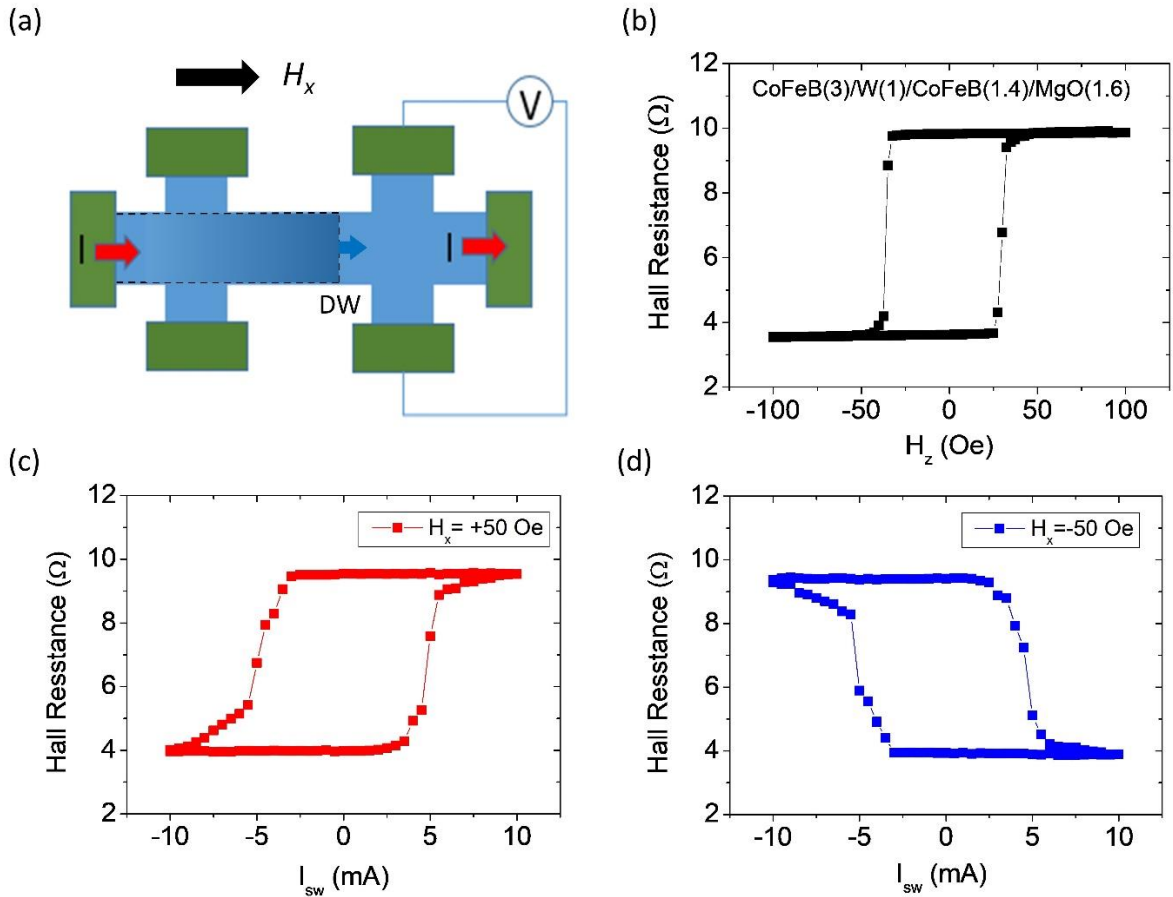

Figure 2. (a) Schematic illustration of electrical measurement setup. DW represents the domain wall location. (b) Representative out-of-plane hysteresis loop obtained through AHE from a CoFeB(3)/W(1)/CoFeB(1.4)/MgO(1.6) sample. Conventional deterministic current-induced SOT switching of a CoFeB/W/CoFeB/MgO sample with a constant supply of in-plane field for (c) $H_x = 50\,\text{Oe}$ and (d) $H_x = -50\,\text{Oe}$.

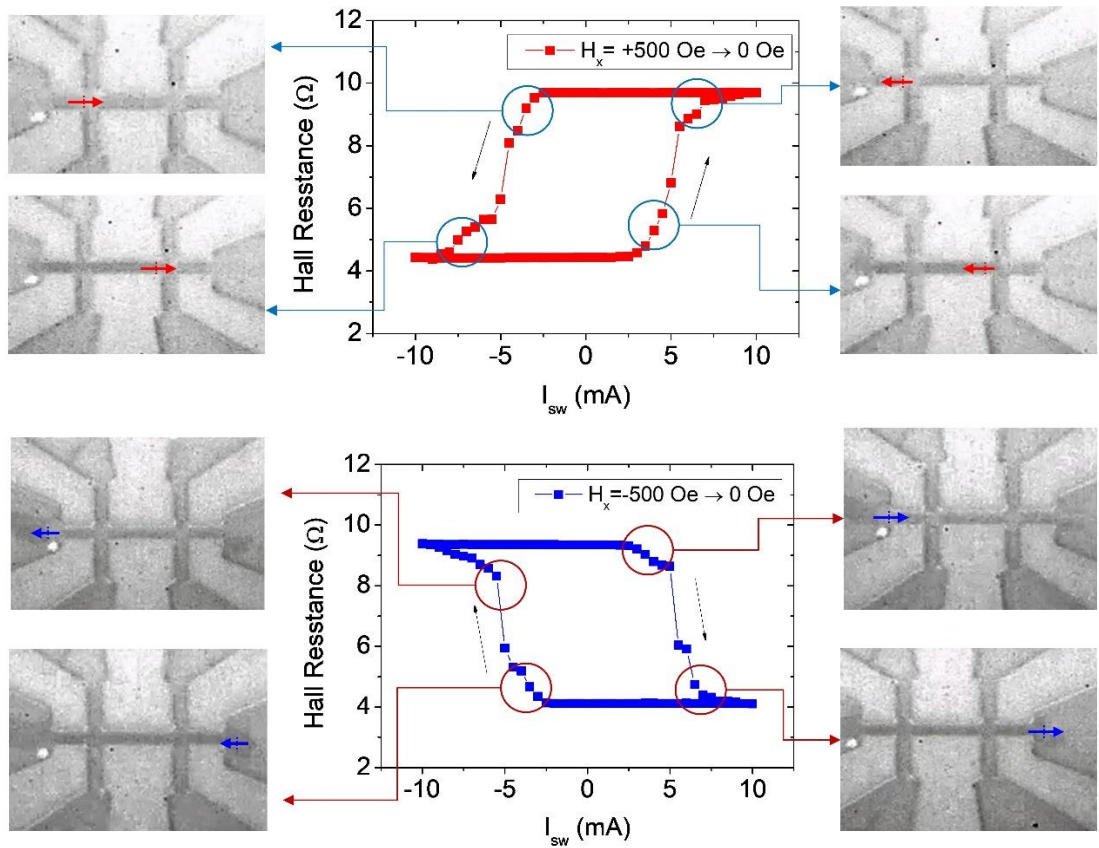

Figure 3. Field-free current-induced SOT magnetization switching and corresponding MOKE images for domain wall motion obtained from a CoFeB(3)/W(1)/CoFeB(1.4)/MgO(2) sample. $H_x = 500\,\text{Oe} \to 0\,\text{Oe}$ indicates that an external field of 500 Oe is first applied then turned off to pre-magnetize the bottom CoFeB(3) magnetization direction. The MOKE images have been edited after imaging to have increased contrast to improve readability.

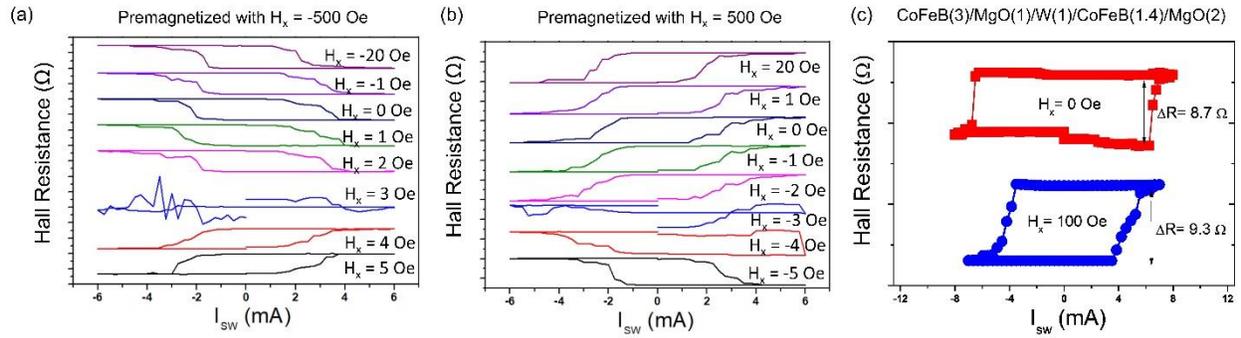

Figure 4. (a) Current-induced SOT magnetization switching at different in-plane field strengths after application of $H_x = -500\,\text{Oe}$. The $H_x = -20\,\text{Oe}$ loop was measured first and the $H_x = 5\,\text{Oe}$ loop was measured last. (b) Current-induced SOT magnetization switching at different in-plane field strengths after application of $H_x = 500\,\text{Oe}$. (c) Current-induced switching measurement results obtained from a CoFeB(3)/MgO(1)/W(1)/CoFeB(1.4)/MgO(2) device with an additional MgO(1) insulating layer under $H_x = 100\,\text{Oe}$ and $H_x = 0\,\text{Oe}$.

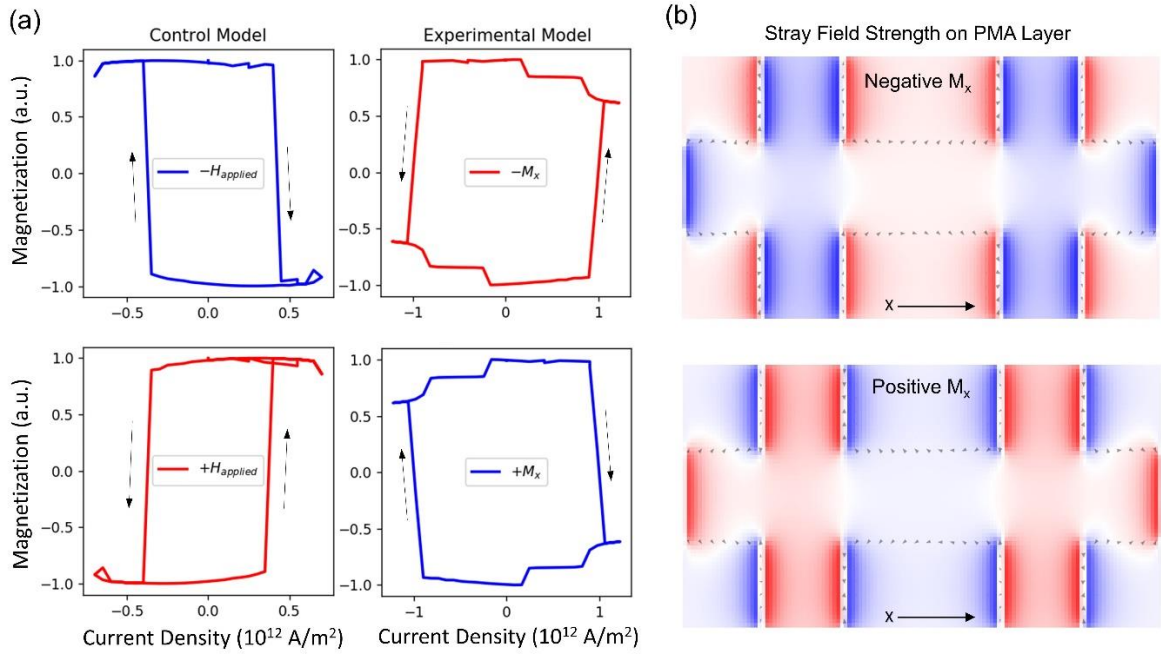

Figure 5. (a) Simulated current-induced SOT switching loops for the control and the experimental models. For the control model, $|H_{applied}| = 100\,\text{Oe}$. (b) The stray field strength in the *x* direction experienced by the top CoFeB layer of the experimental model. Blue and red represent stronger field strengths of opposite direction, and white represents minimal or no field.

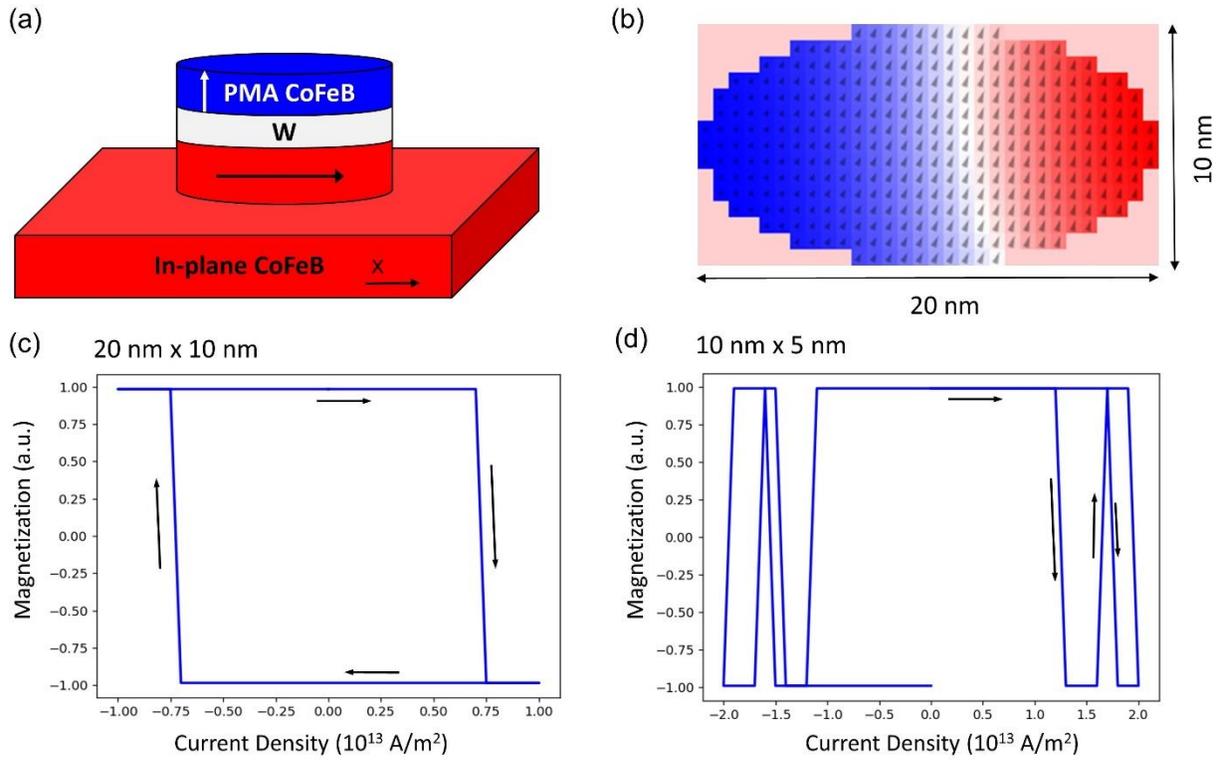

Figure 6. (a) Theoretical design for a SOT-MRAM device with nano-pillar structure with an over-etched in-plane CoFeB base layer providing a stray field to bias the current-induced switching. (b) Simulated current-induced SOT switching through domain wall motion from up (red) to down (blue) of the PMA layer in a 20 nm by 10 nm nano-pillar. (c) Current-induced SOT switching loop of a 20 nm by 10 nm nano-pillar device, using 1 ns pulses with a 10 ns relaxation time. (d) Current-induced SOT switching profile of a 10 nm by 5 nm nano-pillar device.

| Device size $L$ (nm) x $W$ (nm) | Critical current density (A/m$^2$) | Switching mode |
| --- | --- | --- |
| 10 x 5 | 1.25 x 10$^{13}$ | Stochastic |
| 16 x 8 | 8.3 x 10$^{12}$ | Stochastic |
| 20 x 10 | 7.5 x 10$^{12}$ | Deterministic |
| 30 x 15 | 5.35 x 10$^{12}$ | Deterministic |
| 40 x 20 | 4.0 x 10$^{12}$ | Deterministic |
| 60 x 30 | 2.6 x 10$^{12}$ | Deterministic |

Table 1. Simulation results of current-induced SOT switching behaviors for nano-pillar ellipses with different sizes.